\documentclass[11pt,preprint]{aastex}

\def\kms{km\,s$^{-1}$}

\def\deg{$^\circ$}

\def\msunyr{M$_\odot$\,yr$^{-1}$}
\def\mjyb{mJy\,beam$^{-1}$}

\def\cm-2{cm$^{-2}$}
\def\cm-3{cm$^{-3}$}
\def\por{$\times$}

\tighten

\slugcomment{(to be published in AJ, May 2004)}
\shorttitle{VLA 3.6\,cm survey of Galactic WR stars}
\shortauthors{C.E. Cappa, W.M. Goss \& K.A. van der Hucht}

\begin{document}

\title{A Very Large Array 3.6\,cm continuum survey of Galactic Wolf-Rayet 
stars}

\author{C. Cappa\altaffilmark{1}}
\affil{Instituto Argentino de Radioastronom\'{\i}a, CC 5, 1894 Villa Elisa,
       Argentina,                                             \\
       and                                                    \\
       Facultad de Ciencias Astron\'omicas y Geof\'{\i}sicas,
       Universidad Nacional de La Plata,                      \\
       Paseo del Bosque s/n, 1900 La Plata, Argentina}
\email{ccappa@fcaglp.fcaglp.unlp.edu.ar}
\authoraddr{IAR, CC 5, 1894 Villa Elisa, Argentina,
        e-mail: ccappa@fcaglp.fcaglp.unlp.edu.ar}

\author{W.M. Goss}
\affil{National Radio Astronomy Observatory,                  \\
       P.O. Box 0, Socorro, NM\,87801, USA}
\email{mgoss@aoc.nrao.edu}

\and

\author{K.A. van der Hucht}
\affil{SRON National Institute for Space Research             \\
       Sorbonnelaan 2, Utrecht, NL-3584 CA, the Netherlands}
\email{k.a.van.der.hucht@sron.nl}

\altaffiltext{1}{Member of Carrera del Investigador Cient\'{\i}fico,
                 CONICET, Argentina}

\begin{abstract}
We report the results of a survey of radio continuum emission of Galactic
Wolf-Rayet stars north of $\delta$\,=\,$-$46\deg.  The observations were
obtained at 8.46\,GHz (3.6\,cm) using the Very Large Array (VLA), with an
angular resolution of $\sim$\,6\arcsec \por 9\arcsec\ and typical rms noise
of $\sim$\,0.04\,\mjyb.  Our survey of 34 WR stars resulted in 15
definite and  5  probable detections, 13 of these for the first time
at radio wavelengths. 
All detections are unresolved ($\theta$\,$\la$\,5\arcsec).
Time variations in flux are confirmed in the cases of WR\,98a, WR\,104, 
WR\,105, and WR\,125. WR\,79a and WR\,89 are also variable in flux and 
we suspect they are also non-thermal emitters. Thus, of our sample 
20\,-\,30\,\% of the detected stars are non-thermal emiters. Average mass 
loss rates determinations obtained excluding definite and suspected 
non-thermal cases give similar values for WN (all subtypes) and WC5-7 stars 
($\dot{M}$(WN)\,=\,[4\,$\pm$\,3]$\times$10$^{-5}$\,\msunyr\ and 
$\dot{M}$(WC5-7)\,=\,[4\,$\pm$\,2]$\times$10$^{-5}$\,\msunyr), while
a lower value was obtained for WC8-9 stars
($\dot{M}$(WC8-9)\,=\,[2\,$\pm$\,1]$\times$10$^{-5}$\,\msunyr).
Uncertainties in stellar distances largely contribute to the observed
scatter in mass loss rates. 
Upper limits to the mass loss rates were obtained in cases of undetected 
sources or for sources which probably show additional non-thermal emission. 
\end{abstract}



\section{Introduction}

Wolf-Rayet (WR) stars are the evolved descendants of massive O-type
progenitors ($M_{\rm i}$\,$>$\,22\,M$_\odot$) and are considered to be the
immediate precursors of  supernovae Type\,Ib and Type \,Ic (e.g., Maeder
1981; Meynet \& Maeder 2003).  They are characterized by powerful stellar
winds with terminal velocities $v_\infty$\,$\simeq$\,400\,-\,5500\,\kms\
(e.g., Prinja et al.~1990; Eenens \& Williams~1994; Kingsburgh et al.~1995;
for an overview see van der Hucht 2001, table 15, column 14), and mass loss
rates in the range (1\,-\,5){\por}10$^{-5}$\,\msunyr\ (e.g., Abbott et
al.~1986 [hereafter AB86]; Leitherer et al.~1995,~1997 [hereafter LC95,
LC97]; Chapman et al.~1999 [hereafter CL99]).  Due to these properties
WR stars are not only interesting in themselves, but also important species for
chemical enrichment and galactic evolution (e.g., Dray et al.~2003; De
Donder \& Vanbeveren~2003).  For general reviews of the WR phenomenon, see,
e.g., van der Hucht (1992) and Maeder \& Conti (1994).

The strong stellar winds of WR stars reveal themselves in the
electromagnetic spectrum from the UV (P-Cygni profiles of resonance lines)
to the IR/radio range (free-free radiation).  Therefore, an important
diagnostic tool to obtain information about stellar wind properties of
massive stars is the study of the radio continuum emission from their close
environs.  Radio observations provide a reliable method for determining 
stellar mass loss rates if the emission is thermal in origin, since the
measured radio flux density is related to the mass loss rate by an
analytical expression (e.g., Wright \& Barlow 1975; Panagia \& Felli 1975).
This emission is the free-free radiation originating in the ionized wind,
expanding at constant velocity.  It arises from the outer parts of the wind
envelope and is partially optically thick in the radio region, showing positive
spectral indices ($\alpha$\,$\geq$\,+0.6, where
$S_\nu$\,$\propto$\,$\nu^\alpha$).

However, the last decade has seen an increasing number of discoveries of WR
stars with non-thermal radio emission in addition to their free-free 
emission.
Non-thermal emission is thought to be produced by electrons accelerated in
strong shocks in the winds of single stars (White 1985) or in the shocked
wind-wind collision region between massive WR+O binaries (e.g., Williams et
al.~1990; Eichler \& Usov~1993).  Unlike thermal emission, non-thermal
emission has a negative spectral index and can be variable.  Van der Hucht
et al.~(1992) suggested that all cases of WR non-thermal radio emission
correspond to WR+OB colliding wind binaries, which has been corroborated by
Dougherty \& Williams (2000).

Due to their relatively large distances (van der Hucht 2001), the number of
WR radio continuum detections was limited to only about 44 cases (AB86,
LC95, LC97 and CL99).  The VLA survey by AB86, which also quotes previous
observations, contains radio data for 40 WR stars at
$\delta$\,$>$\,$-47$\deg\ with 27 detections using the VLA, with  rms
noise values of 0.06\,-\,0.23\,\mjyb.  They found 77\% probable or definite
thermal wind sources and 12\% probable or definite non-thermal sources.  In
the southern hemisphere, LC95, LC97 and CL99 surveyed 36 WR stars
($\delta$\,$<$\,$-45$\deg) at two frequencies using the ATCA, with rms noise
values of 0.10\,-\,0.15\,\mjyb.  Other WR radio continuum studies have been
performed for WR\,140 by Williams et al.~(1987, 1990, 1994) and White \&
Becker (1995); for WR\,146 by Dougherty et al.~(1996, 2000) and Setia 
Gunawan et al.~(2000); for WR\,147 by
Moran et al.~(1989), Churchwell et al.~(1992), Davis et al.~(1996),
Williams et al.~(1997), Skinner et al.~(1999), Setia Gunawan et al.~(2001),
Watson et al.~(2002) and Dougherty et al.~(2003); for Galactic Center WR 
stars by Lang et al.~(2001) and Lang (2003); for WR\,98a, WR\,104 and 
 WR\,112 by Monnier et al.~(2002); and for
massive stars in Cyg\,OB2 by Setia Gunawan et al.~(2003b).  In these
studies, some 70 out of the $\sim$\,250 known Galactic WR stars (van der
Hucht 2001, 2003) have been investigated for radio continuum emission,
resulting into 44 positive detections.  The ATCA survey also indicates that at
least 40\% of the WR stars with measured radio spectral indices contain
a  non-thermal component at centimeter wavelengths, a percentage that,
coincidentally, is equal to the percentage of WR+OB binaries among the known
Galactic WR stars (van der Hucht~2001).

In this paper we report on the results of a survey of radio continuum
emission from Galactic WR stars at 3.6\,cm with higher sensitivity than
available before.  Our aim is to increase the sample of WR stars with radio
continuum observations in order to derive mass loss rates or upper limits,
and to identify non-thermal cases.  As a first step, we performed
observations at one frequency only.  Consequently, spectral indices of
detected sources can only be determined here using radio data at other
wavelengths from the literature and may be influenced by
variability.  Thus, new radio continuum observations
are important to find new thermal and non-thermal WR radio sources, the
thermal ones providing mass loss rates, crucial for testing existing wind
models, and the non-thermal ones pointing to colliding winds in WR+OB
binaries.


\section{Selected Wolf-Rayet stars}

In order to achieve these goals, we selected 33 WR stars from the catalog
by van der Hucht (2001).  To select the targets, we took into
account all WR stars north of $\delta$\,=$-46$\deg, and selected the ones
with expected free-free flux densities $S_{\rm exp}$\,$>$\,0.075\,mJy.
$S_{\rm exp}$ was determined with the classical expressions by Panagia \&
Felli (1975) and Wright \& Barlow (1975) for the radio emission from an
optically thick stellar wind flowing at constant velocity.  In order to
estimate these values, we adopt for all sources the same mass loss rate
$\dot{M}$\,=\,(1.5\,-\,2.0)\por10$^{-5}$\,\msunyr, and stellar distances
and terminal velocities from van der Hucht (2001).

With these assumptions, we found $S_{\rm exp}$\,$>$\,0.075\,mJy for 64
WR stars.  Only 27 of these stars had previous radio
detections, and radio flux density upper limits were available for 12
stars.  We were then left with 37 stars which either had not been observed
yet or had not been detected.  We selected 26 of these stars as
our targets, listed in Table 1.  In our sample we
included some WR stars previously detected: WR\,79a (HD\,152804, detected
with the VLA by Bieging et al.~1989), WR\,81, WR\,89,
WR\,98a, WR\,104, WR\,105, and WR\,125.  The detection of WR\,81
by AB86 was doubtful.  WR\,98a, WR\,104, WR\,105 and WR\,125 are known
non-thermal emitters and, thus, may vary in flux density.  

\begin{center}
-------------- \\
TABLE 1        \\
--------------
\end{center}

In order to find new non-thermal candidates, we included binaries and
suspected binaries listed by van der Hucht et al.~(2001). About 70\% of 
the stars are closer than 3.0\,kpc, while no stars
beyond 5.3\,kpc are included.


\section{VLA observations}

We performed radio continuum observations at 8.46\,GHz with the VLA in the
direction of the selected WR stars.  Thirty-three fields were observed on
15 September 2001 (during the move from the C- to the DnC-array), 5 and 8
October 2001 (in the DnC-array) and 12 November 2001 (in the D-array).  
With the exception of WR\,89, the field centers coincide with the 
stellar positions. The
bandwidth was 50\,MHz.  The on-source observing time was about 25\,minutes
for each field to achieve a rms noise of $\sim$0.04\,\mjyb.  The
synthesized beam is about 6\farcs 1$\times$9\farcs0 in all cases.

The data were edited, calibrated and imaged following a standard way using
{\sc aips} tasks.  After editing, the data were calibrated both
in amplitude and phase.  The primary flux density calibrators were 3C\,48
and 3C\,286. We used phase referencing taking into account a selection of 
nearby point sources with precision positions (0\farcs 1 or better).
The secondary calibrators were 0244+624, 0747--331, 0828--375, 1001--446,
1604--446, 1744--312, 1751--253, 1820--254, 1832--105, 1851+005, 1924+156,
2015+371, 2230+697 and 2250+558. 
The fields corresponding to WR\,4, WR\,96 and 
WR\,142 were also self-calibrated in phase, since peak flux densities 
within the primary beam exceed 10\,\mjyb.  
Primary beam correction was applied to the images
corresponding to WR\,89 and WR\,105. WR\,89 is in the same field as WR\,87, 
1\farcm 9 far from the field center. The primary beam correction factor for
WR\,89 is 1.5. In the case of WR\,105, two additional point-like sources,
with no relation to the WR star, are detected close to the center 
of the field. Diffuse emission from extended sources
detected in the fields of WR\,87, WR\,104, WR\,106, WR\,113, WR\,114,
WR\,121, WR\,125, WR\,142, WR\,143 and WR\,153ab was removed by applying an
{\sc uv}-range $\geq$\,5K$\lambda$ (rejecting spatial scales
$\simeq$\,40\arcsec).


\section{Observed radio emission}

We report 15 definite  and 5 probable detections at 3.6\,cm 
out of the 34 observed WR stars. For 10 out of the 15 definite detections
and 3 out of the 5 probable detections this represents a first radio 
detection, while four definite detections had been previously identified 
at 3.6\,cm and one had been detected previously at a different frequency.  
All 20 detections are unresolved.  The images of the certain 
and probable detections are displayed in Figs.\,1 to 20. The 
synthesized beam of each field is shown in a corner of the contour plots.

Radio detections with signal-to-noise S\,/\,N\,$\geq$\,5, as derived from 
the Gaussian fitting, and close agreement between radio and optical 
positions were classified as definite detections, while probable detections 
correspond to sources with S\,/\,N\,$\approx$\,4 (WR\,81, WR\,95, WR\,114 
and WR\,155) or with relatively large difference between the optical and 
the radio positions (WR\,79a and WR\,95). 
Detections classified as probable need further confirmation. 

\begin{center}
--------------  \\
FIGURES 1 - 20 \\
--------------
\end{center}

Flux densities and radio positions were obtained by fitting a 2-D Gaussian
with the {\sc aips-jmfit} tool.  In most of the detections, the uncertainty
in the radio position is less than 1\arcsec.  Except for WR\,79a and
WR\,95, the optical and radio positions agree within 2\arcsec.  The flux
density uncertainties quoted in the table are equal to the rms noise
(1$\sigma$) of the images.  
Flux density uncertainties were derived from the images by taking
into account a relatively large  region near the detected sources. 
For the 14 undetected sources, the upper
limits in flux density correspond to 3\,$\sigma$.  Typically, the
1\,$\sigma$ level is in the range 0.025\,-\,0.084\,\mjyb.  In the cases of
WR\,79a and WR\,89, the 1\,$\sigma$ level is 0.13\,\mjyb.  
The large rms noise associated with the
detection of WR\,89 is due to the fact that this star is displaced about 
1\farcm9 from the field center. The typical uncertainty in the radio
position is 0.5\arcsec.

Table\,2 lists the observed 3.6\,cm flux densities and radio coordinates of
the  definite and probable detections,  along with the optical 
coordinates from the catalog of van der Hucht (2001).  Non-detected 
sources are also indicated at the botton of the table. The latter have 
typical uncertainties of about 1\arcsec\ (Leitherer et al. 1997). 

\begin{center}
-------------- \\
TABLE 2       \\
--------------
\end{center}

In Table\,3, we compare our 3.6\,cm detections with previously determined 
flux densities or upper limits at six radio wavelengths. The table also
gives spectral indices $\alpha$ when calculable.

Since we performed observations at one frequency only, spectral indices
can not be derived from our data. Thus, the nature of the emission was 
inferred by comparing the observed fluxes with earlier determinations, 
allowing us to identify variable sources. Variability in flux density, 
which is a hint for the presence of non-thermal emission, is also 
indicated in Table 3.
 
\begin{center}
-------------- \\
TABLE 3        \\
--------------
\end{center}


\section{Comments on individual stars}

\noindent
{\bf WR\,79a}, WN9ha (Fig.\,4).
This object is a known visual binary with $\Delta\phi$\,=\,6\farcs9 (Mason
et al.~1998; van der Hucht 2001, table\,22).  The radio source extends to
the W.  The earliest flux density determinations by BA89 (VLA) indicated
thermal emission with a spectral index $\alpha$\,=\,$+$0.8.  Recent ATCA
observations (November 2000) by Setia Gunawan et al.~(2003a) show a flat
non-thermal energy distribution with spectral index
$\alpha$\,=\,$-$0.0\,$\pm$\,0.1.  Our 3.6\,cm observation, less than a year
later, is consistent with the ATCA observation and, with respect to BA89
observation, with a variable non-thermal component.  

\noindent
{\bf WR\,81}, WC9 (Fig.\,5).
Our VLA image shows a source with a complex structure coincident with the
optical position of the star.  We identified the source to the N as the WR
star because of its proximity (2\farcs 5) to the optical position.  
The radio source detected by AB86 and identified by these authors as
the WR star (see their Fig.\,1) is 4\farcs 6 south of the optical position
and closer to the southern extension of the radio source detected in
the new image. 
A spectral index cannot be derived.

\noindent
{\bf WR\,89}, WN8h+OB (Fig.\,7).
This source was detected in the image corresponding to WR\,87, 1\farcm9
displaced from the field center. Previous flux density determinations by
LC95 were consistent with a thermal wind.  Judging  from our
observation, the 3.6\,cm flux density dropped some 30\,\% in recent years.
WR\,89 is associated with the open cluster Havlen-Moffat\,No.\,1
(C1715$-$387, V\'azquez \& Baume 2001).

\noindent
{\bf WR\,98a}, WC8-9vd+? (Fig.\,9).
In the infrared this object shows a expanding dust spiral, rotating with a
period $P$\,=\,565\,$\pm$\,50\,d (Monnier et al.~1999), which is
considered to be the binary period.  Monnier et al.~(2002) observed this
object with the VLA in 1999.7 and 2000.2 and found the source to be 
non-thermal.  Our 3.6\,cm flux density determination is $\sim$\,20\,\% 
below their result. Apparently, the non-thermal source is variable.  

\noindent
{\bf WR\,104}, WC9d+B0.5V (Fig.\,12).
In the infrared this object shows a expanding dust spiral, rotating with a
period  $P$\,=\,243\,$\pm$\,3 d (Tuthill et al.~1999), which is
considered to be the binary period.  Monnier et al.~(2002) found the source
to be non-thermal.  Our 3.6\,cm flux density determination is
$\sim$\,40\,\% below theirs, thus, the non-thermal source is apparently 
variable. 

\noindent
{\bf WR\,105}, WN9h (Fig.\,13).
This star shows variable non-thermal flux densities.  Our 3.6\,cm flux
density is $\sim$\,50\,\% larger than the one measured by LC97.  The double
radio source to the S of the optical position of WR\,105 was also detected
by AB86 (see their fig.\,1).  These radio sources were not detected in the
NVSS (Condon et al. 1998), nor were optical counterparts found.  
Zoonematkermani et al.~(1990,
radio continuum at 20\,cm) detected the double source with flux densities
of 22\,mJy and 16\,mJy, while from our VLA data we estimate
6.4\,$\pm$\,0.2\,mJy and 5.4\,$\pm$\,0.2\,mJy.  These non-thermal sources
have spectral indices $\alpha$ of $-$0.7 and $-$0.6, respectively.

\noindent
{\bf WR\,125}, WC7ed+O9III (Fig.\,17).
The AB86 1982 observations at 6.3\,cm and 2.0\,cm showed a non-thermal 
radio source  with $\alpha$\,=\,$-$0.5. Our 2001 3.6\,cm flux density 
observation agrees with the AB86 data for a non-thermal source with
that spectral index. The source  is variable in flux and turned 
from non-thermal to thermal in WH92 (see Table 3). We suggest here that 
the emission is again non-thermal because of the similarity in flux 
density levels between the AB86 and the present data.
The detection of non-thermal emission in 1982 and in 2001 suggests a
period of around 20 years.
On the other hand, the last IR excess of the WC+O binary WR\,125 started 
in 1991 (WH92), continued till 1993 and then faded away. IR monitoring 
performed during the last 18 years showed only one period of infrared excess 
(Williams 2002), implying an orbital period of at least 18 years.
WR\,125 has much in common with the archetype WR colliding wind
binary WR\,140 (HD\,193793, WC7pd+O4-5, Williams et al.~1990), which has
a thermal to non-thermal radio light curve with a period of 7.94\,yr
(Williams et al.~1994).  
If WR\,125 is an analog of WR\,140, then its 1992 IR excess occurred at 
periastron passage in an eccentric orbit, and an IR excess will 
happen again in $\sim$2010\,-\,2012.

\noindent
{\bf WR\,133}, WN5+O9I (Fig.\,18).
We detect a double point-like source (separation $\sim$\,10\arcsec) with
the southern component coincident with the optical position of the WN5+O9I
binary (Fig.\,18, upper panel), which is a known visual binary 
($\Delta\phi$\,=\,5\farcs4, Hartkopf et al.~1999; van der Hucht 2001, 
table 22).  The source to the NE has a flux density of 
$S_{\rm 3.6cm}$\,=\,0.39\,$\pm$\,0.03 \mjyb.  The lower panel of 
Fig.\,18 displays an overlay of the radio continuum emission and the DSS2
red image,
suggesting the presence of an additional star NE of the WR system,
coincident in position with the northern radio source.  Pollock et
al.~(1995) found a {\sl ROSAT} X-ray excess for WR\,133, indicative for a
colliding wind binary.

Finally, we note that a radio source with a flux density $S_{3.6cm}$ = 
0.15\,$\pm$\,0.04 mJy is detected 2\farcs 5 far from the optical position 
of WR\,4. We believe that this star has probably been barely detected, but, 
because of the low S\,/\,N ratio and the difference between the optical 
and radio positions (2.1$\sigma$), we have listed this star as non-detected. 

In summary, we note that WR\,79a, WR\,89, WR\,98a, WR\,104, WR\,105 and
WR\,125 are variable, which can be explained by the presence of a 
non-thermal component. This implies that our observations of these six 
stars cannot really be used for mass loss rate determinations, or will 
yield only upper limits.  
Of the above six cases all three WCd objects (WR\,98a, WR\,104 and WR\,125)
turn out to be variable non-thermal radio sources, suggesting a
large number of non-thermal cases between these objects. Additional 
observations are necessary to investigate the nature of the other three 
WCd stars in our sample (WR\,95, WR\,103 and WR\,113) since previous 
observations only provided  upper limits to  the flux densities. 


\section{Radio mass loss rates}

Following Panagia \& Felli (1975) and Wright \& Barlow (1975), the flux
density due to thermal {\it Bremsstrahlung} in an ionized expanding stellar
wind can be expressed as

\[S_{\nu} = 2.32 \times 10^4 (\frac{\dot{M}Z}{v_\infty \mu})^{4/3} (\frac{\gamma g_{\rm ff} \nu}{d^3})^{2/3} \hspace{2cm}(1), \]

\noindent
where $S_{\nu}$ is the flux density in mJy, $\dot{M}$ is the mass loss
rates in \msunyr, $v_\infty$ is the terminal velocity in \kms, $\nu$ is the
frequency in Hz and $d$ is the distance in kpc.  $\mu$, $Z$ and $\gamma$
are the mean molecular weight, the rms ionic charge and the mean number
of electrons per ion, respectively.  This expression corresponds to a
spherically symmetric, stationary,  isothermal wind  flowing at
constant velocity.  The free-free Gaunt factor $g_{\rm ff}$ can be
approximated, following Leitherer \& Robert (1991) as

\[g_{\rm ff} = 9.77 (1 + 0.13 log {\frac{T_{\rm e}^{3/2}}{Z \nu}}) \hspace{2cm} (2). \]

\noindent
In this expression, $T_{\rm e}$ is the electron temperature of the wind in K.

Based on Equations (1) and (2), we have derived mass loss rates
$\dot{M}$ corresponding to the observed flux densities at 3.6\,cm,
assuming that the observed radio emission is due to free-free emission.  In
the case of non-thermal contributions, our values represent upper limits to
the true mass loss rate.  All non-detections are indicated as upper limits
to $\dot{M}$.  Following LC97, we adopted $T_{\rm e}$\,=\,10$^4$\,K in
expression (2).  As pointed out by these authors, this temperature has a
very minor effect on the $g_{\rm ff}$.  Values for the mean molecular
weight, the rms ionic charge and the mean number of electrons per ion
were taken from LC97.  The values adopted for the WO star in
our sample, WR\,142, are those for the WC4 subtype (Crowther et al.~1998).

Table\,4 lists the derived mass loss rates $\dot{M}$ and upper limits.
Uncertainties in these values come from several sources.  Errors in mass
loss rate determinations were estimated following eq. (5) in LC97.  We
adopted similar criteria as LC97 to estimate the uncertainty in each
parameter.  For the error in the distances we assumed 20\% in case of
cluster/associations distances and 50\% in case of field stars.
The logarithmic uncertainty is then $\pm$0.09 and $\pm$0.25,
respectively.
Errors in $g_{\rm ff}$ are about 10\% (0.05 in logarithmic scale).  We
adopted uncertainties in wind terminal velocities of 10\% (i.e., a
logarithmic error of $\pm$0.04).  Finally, logarithmic errors in $Z$,
$\gamma$ and $\mu$ were adopted as $\pm$0.08.  Typical logarithmic errors in
log\,$\dot{M}$ are in the range $\pm$\,0.20 to $\pm$\,0.24 in case of
cluster/association distances and $\pm$\,0.40 in case of field stars.
With the quoted uncertainties for the stellar distances  and other wind
parameters, uncertainties in mass-loss rates determinations are
dominated by errors in distances, while uncertainties in flux densities 
have a minor impact.  

\begin{center}
-------------- \\
TABLE 4        \\
--------------
\end{center}

Other sources of error are non-thermal contributions to the flux at 3.6\,cm.
Inhomogeneities or clumping in the stellar winds lead to over estimates
of the final values as emphasized by, e.g., Contreras et al.~(1996) and
Morris et al.~(2000).  The latter found volume filling factors of the
range 0.04\,-\,0.25, leading to mass-loss rates a factor of
2.5\,-\,5 lower than expected for smooth, homogeneous winds. These 
effects have not been considered here.


\section{Discussion}

In order to compare mass loss rates derived for different WR subtypes from
available radio surveys, we plotted the ATCA and VLA mass loss rates {\it
versus} WR subtype in Fig.\,21. The ATCA surveys have been performed by
LC95, LC97 and CL99.  Previous VLA results were published by Willis 
(1991) (who updated AB86 results by taking into account new criteria in 
terminal velocities and in the chemical composition of the region where radio
emission originates) and by Monnier et al.~(2002).  

\begin{center}
-------------- \\
FIGURE 21      \\
--------------
\end{center}

Monnier et al. observed WR\,98a and WR\,104 at several frequencies
and found a non-thermal component in addition to the
thermal one. They separated both components and derived a mass-loss rate
from the best-fitting value to the thermal emission. The $\dot{M}$-values 
they derived are 0.8\por 10$^{-5}$ \msunyr\ for WR\,98a and  
0.5\por 10$^{-5}$ \msunyr\ for WR\,104.

The upper panel of Fig.\,21 displays the mass loss rate 
($\log\ \dot{M}$) as a function of the spectral WR subtype for the stars 
included in our VLA survey. {\it Bona-fide} mass loss rate estimates are 
indicated by filled circles, while open circles indicate upper limits 
both for the variable detected sources and  the undetected
sources. Uncertainties in the derived values are also indicated.
Note that we have plotted the mass loss rates derived by Monnier et al. 
instead of our estimates listed in Table 4.

The lower panel of Fig.\,21 shows all radio mass loss rates 
from the ATCA and VLA samples. Although more than one mass loss rate 
determination is available for some of the stars, we plotted
only one value for each star from (in
order of preference) Monnier et al.~(2002), new VLA data, ATCA data and
previous VLA surveys.  In the cases of non-thermal sources listed by
LC97 and CL99, the $\dot{M}$-values plotted are upper
limits. As for the upper panel, open circles indicate upper limits
both for undetected sources and variable detected sources. 

Excluding definite and suspected non-thermal cases (WR\,79a, 
WR\,89 and WR\,105) the 
average mass loss rate for the WN stars detected in our survey is
$\dot{M}$(WN)\,=\,[4.3\,$\pm$\,2.5]$\times$10$^{-5}$\,\msunyr, equal to the
average WN mass loss rate in ATCA and previous VLA data sets of
$\dot{M}$(WN)\,=\,4$\times$10$^{-5}$\,\msunyr.  

Mass loss rates determined from emission lines studies by Nugis \& Lamers
(2000)  differ from values derived from radio continuum data:
in some cases (WR\,104 and WR\,113), $\dot{M}$ derived from emission lines 
are factors of two smaller than our radio estimates, while in 
other cases (WR\,81, WR\,95 and WR\,103) factors of two larger than our
results. Clumping and/or an ionization or temperature structure 
different from our assumptions may account for the differences.   

We have thirteen WC8-9 stars included in our sample and detected 
seven  of them.  Four of these stars had no radio detections yet 
(WR\,88, WR\,95, WR\,103 and WR\,113), thus increasing to 12 the number 
of WC8-9 stars with radio detections.  Most of these stars have heated 
circumstellar dust shells (see Williams 1995).  WR\,95, WR\,103  and 
WR\,113 are known to have persistent dust shells (see Table\,1).  
The distribution of these stars in the diagram displays the same trend 
as in fig.\,17 of LC97, with relatively low mass loss rates.  
The average mass loss rate for WC8-9 stars detected in our sample, excluding 
definite non-thermal cases (WR\,98a and WR\,104)  is
$\dot{M}$(WC8-9)\,=\,[1.7\,$\pm$\,1.0]$\times$10$^{-5}$\,\msunyr.
The average mass loss rate for WC5-7 stars, obtained from  
{\it bona-fide} values from the ATCA (5 stars) and the previous (2 
stars) and present (1 star) VLA samples, is 
$\dot{M}$(WC5-7)\,=\,[4.4\,$\pm$\,1.6]$\times$10$^{-5}$\,\msunyr, 
higher than the value derived for WC8-9 stars.
Note, on the other hand, that average values for WN and WC5-7 stars are 
equal.
Uncertainties in stellar distances clearly contribute to the observed 
scatter of mass loss rates. 

The presence of dust in the wind envelopes might decrease the escaping UV
photon flux, lowering the thermal radio emission, as suggested by Monnier
et al.~(2002), and thus the derived mass loss rates.  LC97 suggest that
this may explain the low mass loss rates obtained for WC9 stars from radio
data.  Note that the mass loss rates for the WC8 star WR\,98a,
which also displays a dust persistent shell, is also relatively low (see
Fig.\,21, lower panel).

The percentage of non-thermal emitters between the detected sources in our 
sample is 20\,-\,30\,\%. A similar value is derived for the 1986 
VLA survey (which includes also previous results). Our percentage is lower
than the one derived from ATCA data (40\,\%).
 
The case of WR\,89 (suggested to be a binary by van der Hucht 2001), 
which shows time variations in flux density (see Table\,3), hints to
additional non-thermal emission (the radio spectrum was thermal at the time
of the ATCA observations). WR\,79a also appears to be variable in flux,
suggesting the presence of a non-thermal component. However, a variable 
thermal component can not be discarded (e.g., Watson et al.~2002). 
It would be desirable to perform multifrequency radio observations of 
WR\,79a and WR\,89 to look for variability and to analyze the nature of the
flux variations.

Finally, we would like to comment on the undetected stars.  We repeat here
that uncertainties in the expected flux densities $S_{\rm exp}$ are quite
large, arising mainly from uncertainties in distances of WR field stars and
in the expected range of mass loss rates
(1\,-\,5$\times$10$^{-5}$\,\msunyr).  $S_{\rm exp}$ values for WR\,79b,
WR\,92, WR\,96 and WR\,119 are 0.12\,-\,0.26, 0.09\,-\,0.14, 0.10\,-\,0.16
and 0.11\,-\,0.16\,mJy, respectively, which are below our detection limit
(see Table 2).
$S_{\rm exp}$ values were estimated adopting the stellar parameters
of Tables 1 and 4, and mass loss rates in the range
(1.5\,-\,2.0)\por 10$^{-5}$ \msunyr. For the cases of WR\,80
and WR\,121, however, $S_{\rm exp}$\,=\,0.5\,-\,0.8 and 0.4\,-\,0.6 mJy,
respectively. The derived upper limit is 0.17\,mJy (3$\sigma$) for
both stars. Although a mass loss rate lower than
typical can not be discarded, the objects may be at larger distances than
adopted here, thus lowering the expected flux density.  


\section{Conclusions and suggestions for future work}

We performed a survey of 34 Galactic Wolf-Rayet stars at
3.6\,cm  within the declination range of the Very Large Array. 
We report 15 definite and 5 probable detections. Of these 20 sources, 
13 were detected for the first time at radio frequencies.

We confirm time variations in flux of WR\,98a, WR\,104, WR\,105, and
WR\,125, which support the presence of a non-thermal component.  
For WR\,125 (WC7ed+O9III) we suggest a binary period lower  than 20\,yr.  
WR\,79a (HD\,152408, WN9ha) and WR\,89 are also variable in flux and 
we suspect they are also non-thermal emitters. 
Non-thermal radiation is indicative for wind-wind 
collision in WR+OB binaries.  Thus, of our sample 20\,-\,30\,\% of the 
detected stars are non-thermal radio sources.

Averages of mass loss rate determinations yield similar values for WN 
(all subtypes) and WC5-7 stars 
($\dot{M}$(WN)\,=\,[4\,$\pm$\,3]$\times$10$^{-5}$\,\msunyr\ 
and $\dot{M}$(WC5-7)\,=\,[4\,$\pm$\,2]$\times$10$^{-5}$\,\msunyr), while
a slightly lower value was derived for WC8-9 stars
($\dot{M}$(WC8-9)\,=\,[2\,$\pm$\,1]$\times$10$^{-5}$\,\msunyr).

Future observations at several radio frequencies should monitor the objects
WR\,79a, WR\,89, WR\,98a, WR\,104 and WR\,125 during long time periods. 
These observations can help to determine the presence of non-thermal 
contributions to the flux density  as a function of time.
Radio observations at low and high frequencies of the sources first 
detected will be useful to investigate the nature of the radio emission.

\acknowledgements
{\it Acknowledgements.}
We thank the anonymous referee for very helpful comments and suggestions
that help to improve the presentation of the paper.
C.C. is grateful to the kind hospitality during her stay at AOC, NRAO in
Socorro, New Mexico.  The National Radio Astronomy Observatory is a
facility of the National Science Foundation operated under cooperative
agreement by Associated Universities, Inc.  This research was partially
supported by Facultad de Ciencias Astron\'omicas y Geof\'{\i}sicas,
Universidad Nacional de La Plata; project 11/G049 (UNLP); and CONICET
project PIP 607/98, Argentina.  C.C.  acknowledges a travel grant from
Fundaci\'on Antorchas, Argentina, through project 13622/10.


\clearpage



\begin{table*}
\caption{Selected Wolf-Rayet stars}
\scriptsize
\vspace{3mm}
\label{tbl-1}
\begin{tabular}{rllllrr}
\hline\hline
          &                        &              &       &                           &              &           \\[-2mm] \multicolumn{1}{c}{WR}         &
other                          &
spectral                       &
\multicolumn{1}{c}{$d$}        &
binary                         &
\multicolumn{1}{c}{$P$}        &
\multicolumn{1}{c}{$v_\infty$} \\
\multicolumn{1}{c}{\#}         &
designation (s)                &
type                           &
\multicolumn{1}{c}{(kpc)}      &
status                         &
\multicolumn{1}{c}{(d)}        &
\multicolumn{1}{c}{(\kms)}     \\
          &                        &              &       &                           &              &           \\[-2mm]
\hline\hline
          &                        &              &       &                           &              &           \\[-2mm]
    4~~~  & HD\,16523, V493\,Per   & WC5+?        & ~2.44 & SB1, no d.e.l.            &         2.4  & 1900~~    \\
    5~~~  & HD\,17638              & WC6          & ~1.91 &                           &              & 2100~~    \\
    8~~~  & HD\,62910              & WN7/WCE+?    & ~3.47 & SB1                       &        38.4  & 1590~~    \\
   12~~~  & Ve5-5                  & WN8h+?       & ~5.04 & SB1, no d.e.l.            &        23.9  & 1100~~    \\
   79a~\, & HD\,152408             & WN9ha        & ~1.99 & VB (6\farcs9)             &              &  935~~    \\
   79b~\, & HD\,152386             & WN9ha        & ~2.90 & VB (0\farcs55)            &              & 1650~~    \\
   80~~~  & Wra\,1581              & WC9d         & ~1.50 &                           &              & 1200~~    \\
   81~~~  & He3-1316               & WC9          & ~1.57 &                           &              &  910~~    \\
   82~~~  & LS\,11                 & WN7(h)       & ~5.25 &                           &         2.0  & 1100~~    \\
   87~~~  & LSS\,4064              & WN7(h)+OB    & ~2.88 & a, d.e.l.                 &              & 1400~~    \\
   88~~~  & Th\'e\,1               & WC9          & ~2.33 &                           &              & 1125~~    \\
   89~~~  & AS\,223                & WN8h+OB      & ~2.88 & a, d.e.l., VB (9\farcs88) &              & 1600~~    \\
   92~~~  & HD\,157451             & WC9          & ~3.84 &                           &              & 1100~~    \\
   95~~~  & He3-1434               & WC9d         & ~2.09 &                           &              & 1100~~    \\
   96~~~  & LSS\,4265              & WC9d         & ~3.58 &                           &              & 1100~~    \\
   98a~\, & IRAS\,17380$-$3031     & WC8-9vd+?    & ~1.9  & CWB                       &       565~~~ &  900$^1$~ \\
  100~~~  & HDE\,318139            & WN7          & ~4.61 &                           &              & 1600~~    \\
  103~~~  & HD\,164270, V4072\,Sgr & WC9d+?       & ~2.21 & SB1, no d.e.l.            &         1.8  & 1100~~    \\
  104~~~  & Ve2-45                 & WC9d+B0.5V   & ~2.3  & SB2, VB (0\farcs975)      &       243.5  & 1220~~    \\
  105~~~  & Ve2-47                 & WN9h         & ~1.58 &                           &              &  700~~    \\
  106~~~  & HDE\,313643            & WC9d         & ~2.32 &                           &              & 1100~~    \\
  113~~~  & HD\,168206, CV\,Ser    & WC8d+O8-9IV  & ~1.79 & SB2                       &        29.7  & 1700~~    \\
  114~~~  & HD\,169010             & WC5+OB?      & ~2.00 & d.e.l.                    &              & 2000~~    \\
  119~~~  & Th\'e\,2               & WC9d         & ~3.31 &                           &              & 1200~~    \\
  120~~~  & Vy1-3                  & WN7          & ~3.56 &                           &              & 1225~~    \\
  121~~~  & AS\,320                & WC9d         & ~1.83 &                           &              & 1100~~    \\
  124~~~  & 209\,BAC, QR\,Sge      & WN8h         & ~3.36 & SB1?, no d.e.l.           &         2.7  &  710~~    \\
  125~~~  & IC14-36, V378\,Vul     & WC7ed+O9III  & ~3.06 & SB2                       & $>$\,6600~~~ & 2900~~    \\
  133~~~  & HD\,190918, V1676\,Cyg & WN5+O9I      & ~2.14 & SB2, VB (5\farcs4)        &       112.4  & 1800~~    \\
  142~~~  & Sand\,5                & WO2          & ~0.95 &                           &              & 5500$^2$~ \\
  143~~~  & HD\,195177             & WC4+OB?      & ~1.07 & d.e.l.                    &              & 2750~~    \\
  153ab   & HD\,211853, GP\,Cep    & WN6/WCE +O6I & ~2.75 &                           &         6.7  & 1785~~    \\
  155~~~  & HD\,214419, CQ\,Cep    & WN6+O9II-Ib  & ~2.75 & SB2                       &         1.6  & 1400~~    \\
  156~~~  & AC\,+60\,38562         & WN8h+OB?     & ~3.56 & d.e.l.                    &         6.5  &  660~~    \\[-2mm]
          &                        &              &       &                           &              &           \\
\hline\hline
\end{tabular}

\vspace{2mm}
Notes: \\
Cols.1 and 2: WR catalogue number and any other designation(s); \\ 
col. 3: spectral classification; col. 4: heliocentric distance; \\
col. 5: the binary status, in case of visual binaries (VB), the angular 
separation is indicated; \\ 
col. 6: period;\\ 
col. 7: terminal wind velocity. \\
a: absorption lines present in optical spectrum; \\
d.e.l.: diluted emission lines                  \\
All parameters and values from van der Hucht~(2001) except: \\
$1$: from Monnier et al.~(2002);                 \\
$2$: from Kingsburgh et al.~(1995).
\end{table*}

\normalsize

\clearpage


\begin{table*}
\caption{Optical positions, radio positions, and flux densities of the observed Wolf-Rayet stars}
\small
\label{tbl-2}
\vspace{3mm}
\begin{tabular}{rllcccl}
\hline\hline
     &             &              &                              &                           &                        &      \\[-2mm]
\multicolumn{1}{c}{WR} & \multicolumn{2}{c}{optical position} & \multicolumn{2}{c}{radio position}  & $S_{\rm 3.6cm}$ & epoch \\
\cline{2-5}
     &             &              &                              &                           &                        &      \\[-2mm]
\multicolumn{1}{c}{\#} & R.A.(J2000) & Dec.(J2000) & R.A.(J2000)         & Dec.(J2000) \\
     &  \multicolumn{1}{c}{h ~m~ s} & \multicolumn{1}{c}{$^\circ$ ~$\arcmin$~ $\arcsec$} & h ~m~ s & $^\circ$ ~$\arcmin$~ $\arcsec$ & (mJy) &  \\
     &             &              &                              &                           &                        &         \\[-2mm]
\hline\hline
     &             &              &                              &                           &                       &          \\[-2mm]
\multicolumn{7}{c}{Definite detections}\\[-2mm]
     &             &              &                              &                           &                        &         \\[-2mm]
   5~~~ & 02 52 11.66 & $+$56 56 07.1 & 02 52 11.62\,$\pm$\,0.05 & $+$56 56 08.2\,$\pm$\,0.4 & 0.20\,$\pm$\,0.03      & 2001.9 \\
   8~~~ & 07 44 58.22 & $-$31 54 29.6 & 07 44 58.26\,$\pm$\,0.02 & $-$31 54 29.7\,$\pm$\,0.2 & 0.46\,$\pm$\,0.03      & 2001.8 \\
  12~~~ & 08 44 47.25 & $-$45 58 55.5 & 08 44 47.33\,$\pm$\,0.03 & $-$45 58 52.5\,$\pm$\,0.7 & 0.51\,$\pm$\,0.06      & 2001.8 \\
  88~~~ & 17 18 49.50 & $-$33 57 39.8 & 17 18 49.67\,$\pm$\,0.05 & $-$33 57 40.9\,$\pm\,$0.6 & 0.26\,$\pm$\,0.05      & 2001.8 \\
  89~~~ & 17 19 00.52 & $-$38 48 51.2 & 17 19 00.49\,$\pm$\,0.03 & $-$38 48 49.4\,$\pm\,$0.4 & 2.0~~$\pm$\,0.1~       & 2001.7 \\
  98a~\,& 17 41 12.9  & $-$30 32 29   & 17 41 13.08\,$\pm$\,0.03 & $-$30 32 30.0\,$\pm\,$0.3 & 0.47\,$\pm$\,0.05      & 2001.8 \\
 100~~~ & 17 42 09.77 & $-$32 33 24.7 & 17 42 09.78\,$\pm$\,0.02 & $-$32 33 24.9\,$\pm\,$0.2 & 0.47\,$\pm$\,0.04      & 2001.8 \\
 103~~~ & 18 01 43.14 & $-$32 42 55.2 & 18 01 43.20\,$\pm$\,0.05 & $-$32 42 55.7\,$\pm\,$0.7 & 0.21\,$\pm$\,0.04      & 2001.7 \\
 104~~~ & 18 02 04.07 & $-$23 37 41.2 & 18 02 04.16\,$\pm$\,0.02 & $-$23 37 41.0\,$\pm\,$0.3 & 0.54\,$\pm$\,0.06      & 2001.8 \\
 105~~~ & 18 02 23.46 & $-$23 34 37.7 & 18 02 23.48\,$\pm$\,0.01 & $-$23 34 37.3\,$\pm\,$0.3 & 5.4\,$\pm$\,0.1      & 2001.9 \\
 113~~~ & 18 19 07.36 & $-$11 37 59.2 & 18 19 07.40\,$\pm$\,0.01 & $-$11 37 58.9\,$\pm\,$0.1 & 0.75\,$\pm$\,0.04      & 2001.8 \\
 120~~~ & 18 41 00.88 & $-$04 26 14.3 & 18 41 00.85\,$\pm$\,0.03 & $-$04 26 14.6\,$\pm\,$0.4 & 0.40\,$\pm$\,0.04      & 2001.9 \\
 125~~~ & 19 28 15.57 & $+$19 33 21.1 & 19 28 15.61\,$\pm$\,0.01 & $+$19 33 21.6\,$\pm\,$0.1 & 1.14\,$\pm$\,0.03      & 2001.9 \\
 133~~~ & 20 05 57.33 & $+$35 47 18.2 & 20 05 57.31\,$\pm$\,0.02 & $+$35 47 18.6\,$\pm\,$0.3 & 0.36\,$\pm$\,0.03      & 2001.9 \\
 156~~~ & 23 00 10.13 & $+$60 55 38.4 & 23 00 10.14\,$\pm$\,0.01 & $+$60 55 38.4\,$\pm\,$0.1 & 1.06\,$\pm$\,0.03      & 2001.9 \\[-2mm]
        &             &               &                          &                      &                             &         \\
\multicolumn{7}{c}{Probable detections}\\[-2mm]
        &             & \\
  79a~\,& 16 54 58.51 & $-$41 09 03.1 & 16 54 58.53\,$\pm$\,0.05 & $-$41 09 07.8\,$\pm$\,0.8 & 0.68\,$\pm$\,0.13      & 2001.7 \\
  81~~~ & 17 02 40.39 & $-$45 59 15.5 & 17 02 40.48\,$\pm$\,0.06 & $-$45 59 12.9\,$\pm\,$1.4 & 0.29\,$\pm$\,0.07      & 2001.8 \\
  95~~~ & 17 36 19.76 & $-$33 26 10.9 & 17 36 19.87\,$\pm$\,0.05 & $-$33 26 13.4\,$\pm\,$0.7 & 0.16\,$\pm$\,0.04      & 2001.7 \\
 114~~~ & 18 23 16.39 & $-$13 43 25.8 & 18 23 16.42\,$\pm$\,0.04 & $-$13 43 25.9\,$\pm\,$0.2 & 0.15\,$\pm$\,0.04      & 2001.8 \\
 155~~~ & 22 36 53.96 & $+$56 54 21.0 & 22 36 53.79\,$\pm$\,0.10 & $+$56 54 21.5\,$\pm\,$0.7 & 0.12\,$\pm$\,0.03      & 2001.9 \\
        &             &               &                          &                      &                             &         \\
\multicolumn{7}{c}{Undetected sources}\\[-2mm]
      &  \\
   4~~~ & 02 41 11.68 & $+$56 43 49.7 &        ---               &          ---              & $<$\,0.15              & 2001.9 \\
  79b~\,& 16 55 06.45 & $-$44 59 21.4 &        ---               &          ---              & $<$\,0.20              & 2001.7 \\
  80~~~ & 16 59 02.2  & $-$45 43 06   &        ---               &          ---              & $<$\,0.17              & 2001.8 \\
  82~~~ & 17 04 04.61 & $-$45 12 15.0 &        ---               &          ---              & $<$\,0.23              & 2001.7 \\
  87~~~ & 17 18 52.89 & $-$38 50 04.5 &        ---               &          ---              & $<$0.24                & 2001.7 \\
  92~~~ & 17 25 23.15 & $-$43 29 31.9 &        ---               &          ---              & $<$\,0.18              & 2001.8 \\
  96~~~ & 17 36 24.2 & $-$32 54 29    &        ---               &          ---              & $<$\,0.16              & 2001.7 \\
 106~~~ & 18 04 43.66 & $-$21 09 30.7 &        ---               &          ---              & $<$\,0.17              & 2001.8 \\
 119~~~ & 18 39 17.91 & $-$10 05 31.1 &        ---               &          ---              & $<$\,0.11              & 2001.7 \\
 121~~~ & 18 44 13.15 & $-$03 47 57.8 &        ---               &          ---              & $<$\,0.17              & 2001.9 \\
 124~~~ & 19 11 30.88 & $+$16 51 38.2 &        ---               &          ---              & $<$\,0.25              & 2001.9 \\
 142~~~ & 20 21 44.36 & $+$37 22 30.3 &        ---               &          ---              & $<$0.90                & 2001.9 \\
 143~~~ & 20 28 22.68 & $+$38 37 18.9 &        ---               &          ---              & $<$\,0.12              & 2001.9 \\
 153ab  & 22 18 45.61 & $+$56 07 33.9 &        ---               &          ---              & $<$\,0.14              & 2001.9 \\
        &             &               &                          &                      &                             &         \\

\hline\hline
\end{tabular}
\end{table*}

\normalsize

\clearpage


\begin{table*}
\begin{center}
\caption{Radio flux densities of earlier detected Wolf-Rayet
         stars, compared with our detections}
\label{tbl-4}
\vspace{3mm}
\scriptsize
\begin{tabular}{r llllll ll}
\hline\hline
        &                   &                  &                   &                   &                   &                   &                       &            \\[-3mm]
\multicolumn{1}{c}{WR} & \multicolumn{6}{c}{present and previous observations (mJy)}                                           & spectral              & references \\
\cline{2-7}
\multicolumn{1}{c}{\#}
        & $S_{\rm 21cm}$    & $S_{\rm 12.5cm}$ & $S_{\rm 6.3cm}$   & $S_{\rm 3.6cm}$   & $S_{\rm 2.0cm}$   & $S_{\rm 1.3cm}$   & index $\alpha$        &            \\
        &                   &                  &                   &                   &                   &                   &                       &            \\[-3mm]
\hline\hline
        &                   &                  &                   &                   &                   &                   &                       &            \\[-2mm]
 79a    &                   &                  & 1.1\,$\pm$\,0.1   &                   & 2.4\,$\pm$\,0.1   &                   & \,~$+$0.8             & BA89       \\[-1mm]
        &                   & 1.0\,$\pm$\,0.1  & 0.8\,$\pm$\,0.1   & 0.9\,$\pm$\,0.1   &                   &                   & \,~$-$0.0             & SC03       \\[-1mm]
        &                   &                  &                   & 0.7\,$\pm$\,0.1   &                   &                   & ~~~~V                 & this study \\[-1mm]
        &                   &                  &                   &                   &                   &                   &                       &            \\[-2mm]
\hline
        &                   &                  &                   &                   &                   &                   &                       &            \\[-2mm]
 81~\,  &                   &                  & 0.3\,$\pm$\,0.08  &                   &                   &                   &                       & AB86       \\[-1mm]
        &                   &                  &                   & 0.3\,$\pm$\,0.1   &                   &                   &                       & this study \\[-1mm]
        &                   &                  &                   &                   &                   &                   &                       &            \\[-2mm]
\hline
        &                   &                  &                   &                   &                   &                   &                       &            \\[-2mm]
 88~\,  &                   &                  & $<$0.42           & $<$0.42           &                   &                   &                       & LC97       \\[-1mm]
        & $<$1.29           & $<$0.57          &                   &                   &                   &                   &                       & CL99       \\[-1mm]
        &                   &                  &                   & 0.26\,$\pm$\,0.05 &                   &                   &                       & this study \\[-1mm]
        &                   &                  &                   &                   &                   &                   &                       &            \\[-2mm]
\hline
        &                   &                  &                   &                   &                   &                   &                       &            \\[-2mm]
 89~\,  &                   &                  & 0.6\,$\pm$\,0.1   &                   &                   &                   &                       & AB86       \\[-1mm]
        &                   &                  & 1.9\,$\pm$\,0.2   & 3.0\,$\pm$\,0.1   &                   &                   & \,~$+$0.8             & LC95       \\[-1mm]
        & $<$1.20           & $<$0.90          &                   &                   &                   &                   &                       & CL99       \\[-1mm]
        &                   &                  &                   & 2.0\,$\pm$\,0.1   &                   &                   & ~~~~V                 & this study \\[-1mm]
        &                   &                  &                   &                   &                   &                   &                       &            \\[-2mm]
\hline
        &                   &                  &                   &                   &                   &                   &                       &            \\[-2mm]
 95~\,  &                   &                  & $<$0.4            &                   &                   &                   &                       & AB86       \\[-1mm]
        &                   &                  & $<$0.45           & $<$0.45           &                   &                   &                       & LC97       \\[-1mm]
        & $<$1.50           & $<$0.66          &                   &                   &                   &                   &                       & CL99       \\[-1mm]
        &                   &                  &                   & 0.16\,$\pm$\,0.04 &                   &                   &                       & this study \\[-1mm]
        &                   &                  &                   &                   &                   &                   &                       &            \\[-2mm]
\hline
        &                   &                  &                   &                   &                   &                   &                       &            \\[-2mm]
 98a    & $<$0.36           &                  & 0.37\,$\pm$\,0.07 & 0.60\,$\pm$\,0.05 & 0.64\,$\pm$\,0.11 & 0.57\,$\pm$\,0.10 & \,~$+$0.3             & MT02       \\[-1mm]
        &                   &                  &                   & 0.47\,$\pm$\,0.05 &                   &                   & ~~~~V                 & this study \\[-1mm]
        &                   &                  &                   &                   &                   &                   &                       &            \\[-2mm]
\hline
        &                   &                  &                   &                   &                   &                   &                       &            \\[-2mm]
103~\,  &                   &                  & $\leq$0.2         &                   &                   &                   &                       & AB86       \\[-1mm]
        &                   &                  & $<$0.42           & $<$0.42           &                   &                   &                       & LC97       \\[-1mm]
        & $<$0.90           & $<$0.42          &                   &                   &                   &                   &                       & CL99       \\[-1mm]
        &                   &                  &                   & 0.21\,$\pm$\,0.04 &                   &                   &                       & this study \\[-1mm]
        &                   &                  &                   &                   &                   &                   &                       &            \\[-2mm]
\hline
        &                   &                  &                   &                   &                   &                   &                       &            \\[-2mm]
104~\,  &                   &                  & $<$0.4            &                   &                   &                   &                       & AB86       \\[-1mm]
        &                   &                  & $<$2.01           & $<$0.39           &                   &                   &                       & LC97       \\[-1mm]
        & $<$1.59           & $<$0.99          &                   &                   &                   &                   &                       & CL99       \\[-1mm]
        & $<$0.30           &                  &                   & 0.87\,$\pm$\,0.06 & 1.02\,$\pm$\,0.12 & 0.94\,$\pm$\,0.10 & \,~$+$0.1             & MT02       \\[-1mm]
        &                   &                  &                   & 0.54\,$\pm$\,0.06 &                   &                   & ~~~~V                 & this study \\[-1mm]
        &                   &                  &                   &                   &                   &                   &                       &            \\[-2mm]
\hline
        &                   &                  &                   &                   &                   &                   &                       &            \\[-2mm]
105~\,  &                   &                  & 3.6\,$\pm$\,0.21  &                   &                   &                   &                       & AB86       \\[-1mm]
        &                   &                  & 4.39\,$\pm$\,0.15 & 3.8\,$\pm$\,0.2   &                   &                   & \,~$-$0.3             & LC97       \\[-1mm]
        & $<$1.17           & $<$0.69          &                   &                   &                   &                   &                       & CL99       \\[-1mm]
        &                   &                  &                   & 5.4\,$\pm$\,0.1   &                   &                   & ~~~~V                 & this study \\[-1mm]
        &                   &                  &                   &                   &                   &                   &                       &            \\[-2mm]
\hline
        &                   &                  &                   &                   &                   &                   &                       &            \\[-2mm]
113~\,  &                   &                  & $\leq$0.4         &                   &                   &                   &                       & BA82       \\[-1mm]
        &                   &                  & $<$0.80           & $<$0.80           &                   &                   &                       & LC97       \\[-1mm]
        & $<$2.25           & $<$0.90          &                   &                   &                   &                   &                       & CL99       \\[-1mm]
        &                   &                  &                   & 0.75\,$\pm$\,0.04 &                   &                   &                       & this study \\[-1mm]
        &                   &                  &                   &                   &                   &                   &                       &            \\[-2mm]
\hline
        &                   &                  &                   &                   &                   &                   &                       &            \\[-2mm]
114~\,  &                   &                  & $<$0.3            &                   &                   &                   &                       & AB86       \\[-1mm]
        &                   &                  & $<$0.45           & $<$0.45           &                   &                   &                       & LC97       \\[-1mm]
        & $<$1.17           & $<$0.54          &                   &                   &                   &                   &                       & CL99       \\[-1mm]
        &                   &                  &                   & 0.15\,$\pm$\,0.03 &                   &                   &                       & this study \\[-1mm]
        &                   &                  &                   &                   &                   &                   &                       &            \\[-2mm]
\hline
        &                   &                  &                   &                   &                   &                   &                       &            \\[-2mm]
125~\,  &                   &                  & 1.5\,$\pm$\,0.09  &                   & 0.8\,$\pm$\,0.09  &                   & \,~$-$0.5             & AB86       \\[-1mm]
        & 1.53\,$\pm$\,0.06 &                  & 1.18\,$\pm$\,0.06 &                   & 0.82\,$\pm$\,0.10 &                   & \,~$-$0.3             & WH92       \\[-1mm]
        &                   &                  & 0.20\,$\pm$\,0.04 &                   &                   &                   & \,~$+$0.7~ V          & WH92       \\[-1mm]
        &                   &                  &                   & 1.14\,$\pm$\,0.03 &                   &                   &                       & this study \\[-1mm]
        &                   &                  &                   &                   &                   &                   &                       &            \\[-2mm]
\hline
        &                   &                  &                   &                   &                   &                   &                       &            \\[-2mm]
133~\,  &                   &                  & $<$0.3            &                   &                   &                   &                       & AB86       \\[-1mm]
        &                   &                  &                   & 0.36\,$\pm$\,0.03 &                   &                   &                       & this study \\[-1mm]
        &                   &                  &                   &                   &                   &                   &                       &            \\[-2mm]
\hline\hline
\end{tabular}
\end{center}

\small
Note:
V: apparently variable.
References:
AB86: Abbott    et al.~(1986);
BA82: Bieging   et al.~(1982);
BA89: Bieging   et al.~(1989);
CL99: Chapman   et al.~(1999);
LC95: Leitherer et al.~(1995);
LC97: Leitherer et al.~(1997);
MT02: Monnier   et al.~(2002);
SC03: Setia Gunawan et al.~(2003a);
WH92: Williams  et al.~(1992).
\end{table*}
\normalsize

\clearpage


\begin{table*}
\caption{Mass loss rates from 3.6\,cm observations of our sample of 34 Wolf-Rayet stars}
\label{tbl-4}
\vspace{3mm}
\begin{tabular}{rlrccrclcl}
\hline\hline
           &       &            &       &     &          &                   &                         &                      &       \\[-2mm]
\multicolumn{1}{c}{WR} & \multicolumn{1}{c}{$d$}   &
\multicolumn{1}{c}{$v_\infty$}  & $\mu$ & $Z$ & $\gamma$ & $S_{\rm 3.6cm}$ & \multicolumn{1}{c}{log\,$\dot{M}$} & $\dot{M}$     & Notes \\
\multicolumn{1}{c}{\#} & \multicolumn{1}{c}{(kpc)} &
\multicolumn{1}{c}{(\kms)}      &       &     &          &    (mJy)          &                         & ($10^{-5}$\,\msunyr) &       \\
           &       &            &       &     &          &                   &                         &                      &       \\[-2mm]
\hline\hline
           &       &            &       &     &          &                   &                         &                      &       \\[-2mm]
   4~~~    & 2.44  & 1900       & 5.1   & 1.2 & 1.0      & $<$\,0.15         & $<$\,$-$4.7             &  $<$\,~2.0           & \\
   5~~~    & 1.91  & 2100       & 4.9   & 1.2 & 1.0      & 0.20\,$\pm$\,0.03 &  ~~\,$-$4.7\,$\pm$\,0.4 &   ~~\,~1.8           &       \\
   8~~~    & 3.47  & 1590       & 1.7   & 1.0 & 1.0      & 0.46\,$\pm$\,0.03 &  ~~\,$-$4.6\,$\pm$\,0.2 &   ~~\,~2.6           &       \\
  12~~~    & 5.04  & 1100       & 2.6   & 1.0 & 1.0      & 0.51\,$\pm$\,0.06 &  ~~\,$-$4.3\,$\pm$\,0.4 &   ~~\,~5.2           &       \\
  79a~\,   & 1.99  &  935       & 2.6   & 1.0 & 1.0      & 0.68\,$\pm$\,0.14 & $<$\,$-$4.9\,$\pm$\,0.2 &  $<$\,~1.4           & NT?    \\
  79b~\,   & 2.9   & 1650       & 2.6   & 1.0 & 1.0      & $<$\,0.20         & $<$\,$-$4.8             &  $<$\,~1.7           &       \\
  80~~~    & 1.5   & 1200       & 4.7   & 1.1 & 1.1      & $<$\,0.17         & $<$\,$-$5.2             &  $<$\,~0.6           &       \\
  81~~~    & 1.57  &  910       & 4.7   & 1.1 & 1.1      & 0.29\,$\pm$\,0.07 &  ~~\,$-$5.1\,$\pm$\,0.4 &   ~~\,~0.8           &    \\
  82~~~    & 5.25  & 1100       & 1.7   & 1.0 & 1.0      & $<$\,0.23         & $<$\,$-$4.7             &  $<$\,~1.9           &       \\
  87~~~    & 2.88  & 1400       & 1.8   & 1.0 & 1.0      & $<$\,0.24         & $<$\,$-$4.9             &  $<$\,~1.2           &       \\
  88~~~    & 2.33  & 1125       & 4.7   & 1.1 & 1.1      & 0.26\,$\pm$\,0.05 &  ~~\,$-$4.8\,$\pm$\,0.4 &   ~~\,~1.6           &       \\
  89~~~    & 2.88  & 1600       & 1.5   & 1.0 & 1.0      & 2.0~~$\pm$\,0.1~  &  ~~\,$-$4.3\,$\pm$\,0.2 &   ~~\,~5.2           & NT?   \\
  92~~~    & 3.84  & 1100       & 4.7   & 1.1 & 1.1      & $<$\,0.18         & $<$\,$-$4.6             &  $<$\,~2.5           &       \\
  95~~~    & 2.09  & 1100       & 4.7   & 1.1 & 1.1      & 0.16\,$\pm$\,0.04 &  ~~\,$-$5.0\,$\pm$\,0.2 &   ~~\,~1.0           &       \\
  96~~~    & 3.58  & 1100       & 4.7   & 1.1 & 1.1      & $<$\,0.16         & $<$\,$-$4.7             &  $<$\,~2.1           &       \\
  98a~\,   & 1.9   & 2000       & 4.7   & 1.1 & 1.1      & 0.47\,$\pm$\,0.05 & $<$\,$-$4.8\,$\pm$\,0.2 &  $<$\,~1.5           & NT    \\
 100~~~    & 4.61  & 1600       & 4.0   & 1.0 & 1.0      & 0.47\,$\pm$\,0.04 &  ~~\,$-$4.0\,$\pm$\,0.4 &   ~~\,~9.5           &       \\
 103~~~    & 2.21  & 1100       & 4.7   & 1.1 & 1.1      & 0.21\,$\pm$\,0.04 &  ~~\,$-$4.9\,$\pm$\,0.4 &   ~~\,~1.2           &       \\
 104~~~    & 2.3   & 1220       & 4.7   & 1.1 & 1.1      & 0.54\,$\pm$\,0.06 & $<$\,$-$4.5\,$\pm$\,0.2 &  $<$\,~2.9           & NT    \\
 105~~~    & 1.58  &  700       & 2.6   & 1.0 & 1.0      & 5.4~~\,$\pm$\,0.1~ & $<$\,$-$4.5\,$\pm$\,0.4 &  $<$\,~3.4           & NT    \\
 106~~~    & 2.32  & 1100       & 4.7   & 1.1 & 1.1      & $<$\,0.17         & $<$\,$-$5.0             &  $<$\,~1.1           &       \\
 113~~~    & 1.79  & 1700       & 4.7   & 1.1 & 1.1      & 0.75\,$\pm$\,0.04 &  ~~\,$-$4.4\,$\pm$\,0.2 &   ~~\,~3.6           &       \\
 114~~~    & 2.0   & 2000       & 4.9   & 1.2 & 1.1      & 0.15\,$\pm$\,0.04 &  ~~\,$-$4.8\,$\pm$\,0.2 &   ~~\,~1.5           &       \\
 119~~~    & 3.31  & 1200       & 4.7   & 1.1 & 1.1      & $<$\,0.11         & $<$\,$-$4.8             &  $<$\,~1.5           &       \\
 120~~~    & 3.56  & 1225       & 4.0   & 1.0 & 1.0      & 0.40\,$\pm$\,0.04 &  ~~\,$-$4.4\,$\pm$\,0.4 &   ~~\,~4.4           &       \\
 121~~~    & 1.83  & 1100       & 4.7   & 1.1 & 1.1      & $<$\,0.17         & $<$\,$-$5.1             &  $<$\,~0.8           &       \\
 124~~~    & 3.36  &  710       & 3.7   & 1.0 & 1.0      & $<$\,0.25         & $<$\,$-$4.9             &  $<$\,~1.3           &       \\
 125~~~    & 3.06  & 2900       & 4.7   & 1.2 & 1.1      & 1.14\,$\pm$\,0.03 & $<$\,$-$3.8\,$\pm$\,0.4 &  $<$\,17.1           & NT    \\
 133~~~    & 2.14  & 1800       & 4.0   & 1.1 & 1.1      & 0.36\,$\pm$\,0.03 &  ~~\,$-$4.6\,$\pm$\,0.2 &   ~~\,~2.8           &    \\
 142~~~    & 0.95  & 5500       & 5.1   & 1.2 & 1.1      & $<$\,0.77         & $<$\,$-$4.3             &  $<$\,~4.5           &       \\
 143~~~    & 1.07  & 2750       & 5.1   & 1.2 & 1.0      & $<$\,0.12         & $<$\,$-$5.2             &  $<$\,~0.7           &       \\
 153ab     & 2.75  & 1785       & 4.0   & 1.0 & 1.0      & $<$\,0.14         & $<$\,$-$4.7             &  $<$\,~2.0           &       \\
 155~~~    & 2.75  & 1400       & 4.0   & 1.0 & 1.0      & 0.12\,$\pm$\,0.03 &  ~~\,$-$4.9\,$\pm$\,0.2 &   ~~\,~1.4           &       \\
 156~~~    & 3.56  &  660       & 3.3   & 1.0 & 1.0      & 1.06\,$\pm$\,0.03 &  ~~\,$-$4.4\,$\pm$\,0.2 &   ~~\,~4.0           &       \\[-2mm]
           &       &            &       &     &          &                   &                         &                      &       \\
\hline\hline
\end{tabular}
\normalsize

\end{table*}

\clearpage


\begin{figure}
\caption{Contour image for WR\,5. The cross marks the optical position of 
the WR star, and the ellipse in one of the corners, the synthesized beam.
We note that the extend of the cross does not represent the uncertainty
in the stellar position. Contour lines are --3 (dashed contour), 3, 4, 5, 
6 and 7$\sigma$ (1$\sigma$ = 0.025 \mjyb).}
\end{figure}

\begin{figure}
\caption{Contour image for WR\,8. 
Contour lines are --3 (dashed contour), 3, 4, 5, 6, 8, 10, 12 and 14$\sigma$ 
(1$\sigma$ = 0.030 \mjyb).}
\end{figure}

\begin{figure}
\caption{Contour image for WR\,12. 
Contour lines are --3 (dashed contour), 3, 4, 5, 6, 7, 8 and 9$\sigma$ 
(1$\sigma$ = 0.055 \mjyb).}
\end{figure}

\begin{figure}
\caption{Contour image for WR\,79a. 
Contour lines are --3 (dashed contour), 3 and 4$\sigma$ 
(1$\sigma$ = 0.125 \mjyb).}
\end{figure}

\begin{figure}
\caption{Contour image for WR\,81. 
Contour lines are --3 (dashed contour) and 3$\sigma$ 
(1$\sigma$ = 0.065 \mjyb).}
\end{figure}

\begin{figure}
\caption{Contour image for WR\,88. 
Contour lines are --3 (dashed contour), 3, 4 and 5$\sigma$ 
(1$\sigma$ = 0.050 \mjyb).}
\end{figure}

\begin{figure}
\caption{Contour image for WR\,89. 
Contour lines are --3 (dashed contour), 3, 4, 5, 6, 7, 9, 11 and 
13$\sigma$ (1$\sigma$ = 0.130 \mjyb).}
\end{figure}

\begin{figure}
\caption{Contour image for WR\,95. 
Contour lines are --3 (dashed contour) and 3$\sigma$ 
(1$\sigma$ = 0.040 \mjyb).}
\end{figure}

\begin{figure}
\caption{Contour image for WR\,98a. 
Contour lines are --3 (dashed contour), 3, 4, 5, 6 and 7$\sigma$ 
(1$\sigma$ = 0.050 \mjyb).}
\end{figure}

\begin{figure}
\caption{Contour image for WR\,100. 
Contour lines are --3 (dashed contour), 3, 4, 5, 6, 7, 8, 10 and 12$\sigma$ 
(1$\sigma$ = 0.040 \mjyb).}
\end{figure}

\begin{figure}
\caption{Contour image for WR\,103. 
Contour lines are --3 (dashed contour), 3 and 4$\sigma$ (1$\sigma$ =
0.040 \mjyb).}
\end{figure}

\clearpage

\begin{figure}
\caption{Contour image for WR\,104. 
Contour lines are --3 (dashed contour), 3, 4, 5, 6 and 7$\sigma$ 
(1$\sigma$ = 0.060 \mjyb).}
\end{figure}

\begin{figure}
\caption{Contour image for WR\,105. 
Contour lines are --3 (dashed contour), 3, 6, 12, 24, 36, 48, and 
60$\sigma$ (1$\sigma$ = 0.080 \mjyb).}
\end{figure}

\begin{figure}
\caption{Contour image for WR\,113. 
Contour lines are --3 (dashed contour), 3, 6, 9, 12, 15 and 17$\sigma$ 
(1$\sigma$ = 0.040 \mjyb).}
\end{figure}

\begin{figure}
\caption{Contour image for WR\,114. 
Contour lines are --3 (dashed contour) and 3$\sigma$ (1$\sigma$ = 
0.040 \mjyb).}
\end{figure}

\begin{figure}
\caption{Contour image for WR\,120. 
Contour lines are --3 (dashed contour), 3, 4, 5, 6, 7, 8, 9 and 10$\sigma$ 
(1$\sigma$ = 0.035 \mjyb).}
\end{figure}

\begin{figure}
\caption{Contour image for WR\,125. 
Contour lines are --3 (dashed contour), 3, 6, 9, 12, 15, 18, 24, 30 and 
35$\sigma$ (1$\sigma$ = 0.030 \mjyb).  }
\end{figure}

\begin{figure}
\caption{Contour image for WR\,133. 
{\it Upper panel: } Contour lines are --3 (dashed contour), 3, 4, 6, 
8, 10, 12 and 14$\sigma$ (1$\sigma$ = 0.025 \mjyb). {\it Lower panel:}
Overlay of the DSS2 red image of WR\,133 (gray scale, in arbitrary units)
and the radio continuum emission at 3.6\,cm.  }
\end{figure}

\begin{figure}
\caption{Contour image for WR\,155. 
Contour lines are --3 (dashed contour), 3 and 4$\sigma$ 
(1$\sigma$ = 0.025 \mjyb).}
\end{figure}

\begin{figure}
\caption{Contour image for WR\,156. 
Contour lines are --3 (dashed contour), 3, 6, 9, 12, 15, 18, 24 and
30$\sigma$ 
(1$\sigma$ = 0.030 \mjyb).}
\end{figure}

\begin{figure}
\caption{WR mass loss rates (log $\dot{M}$, where $\dot{M}$ is in units of 
\msunyr) from radio observations  {\it versus} spectral
subtype.  {\it Upper panel:} Mass loss rates derived from the
present VLA survey.  {\it Bona fide} mass loss rates values corresponding 
to  detected sources are indicated as filled circles. Bar errors are 
included. Open circles indicate upper limits corresponding to both 
detected sources with variable flux density and undetected sources.
{\it Lower panel:} Mass loss rates derived for the VLA and ATCA target 
stars. Open and filled circles have the same meaning as in the upper
panel.
}  
\end{figure}

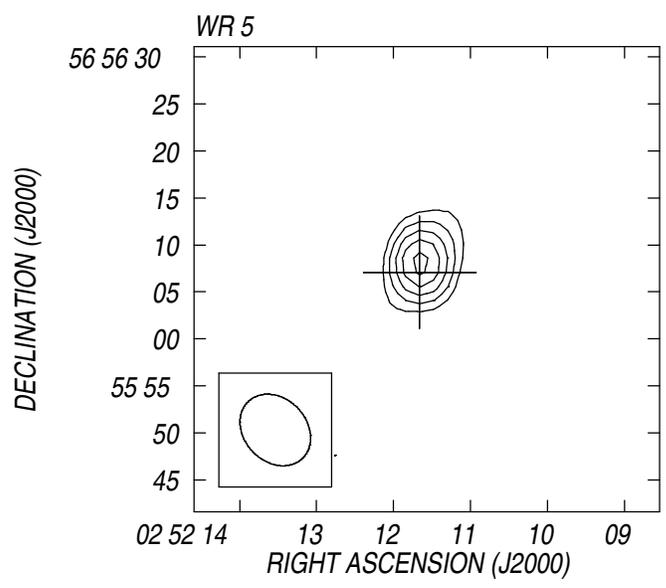
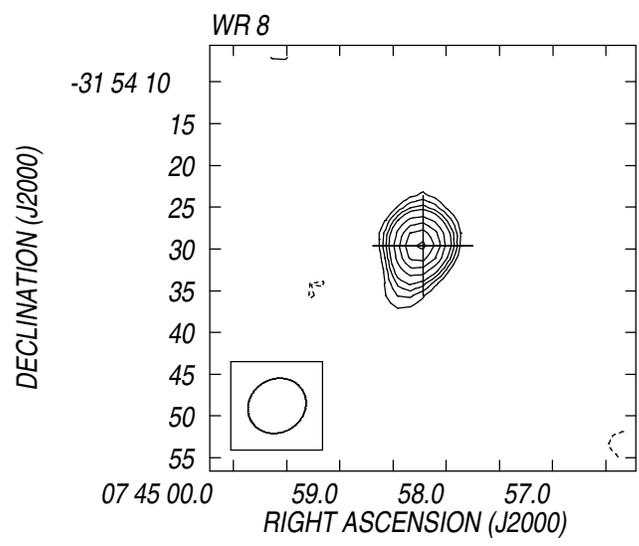
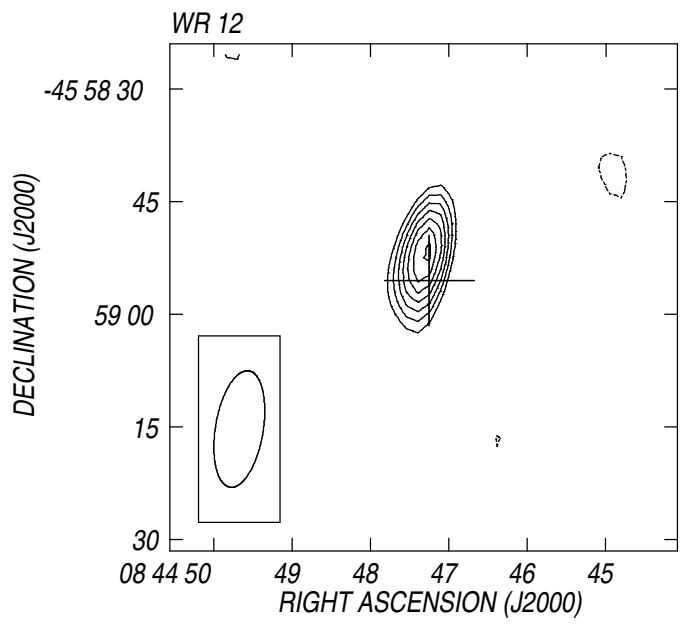
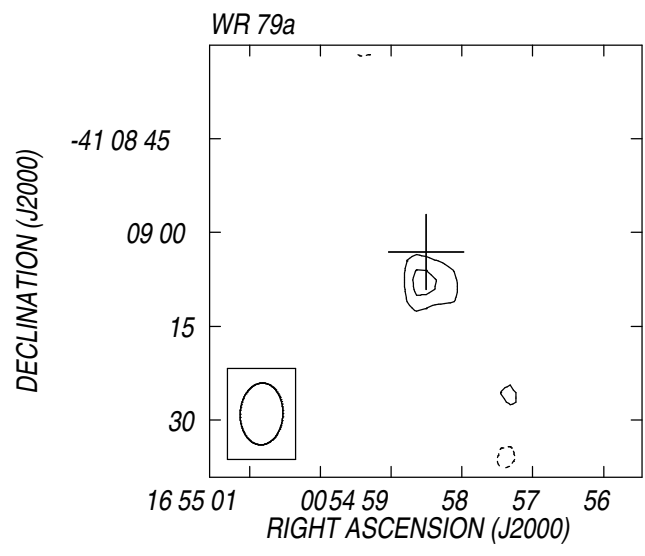
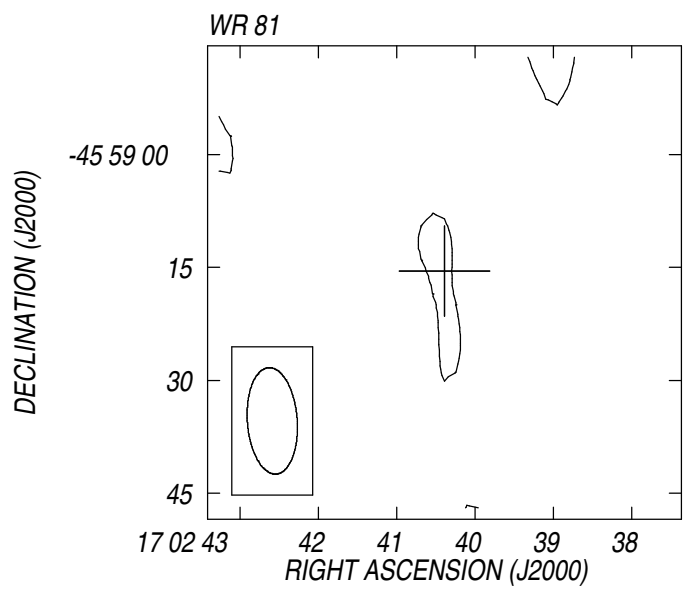
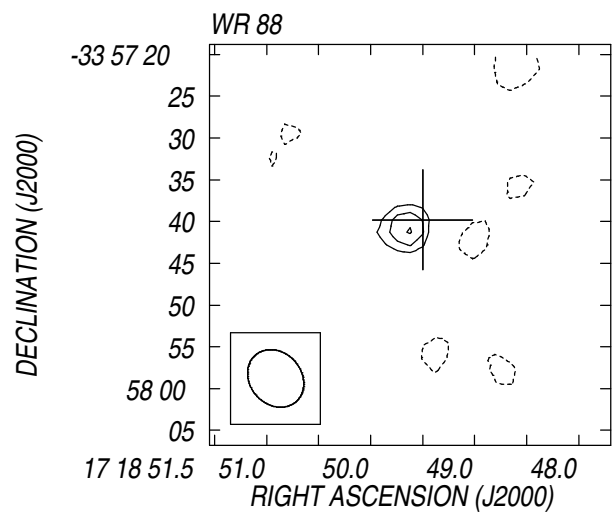
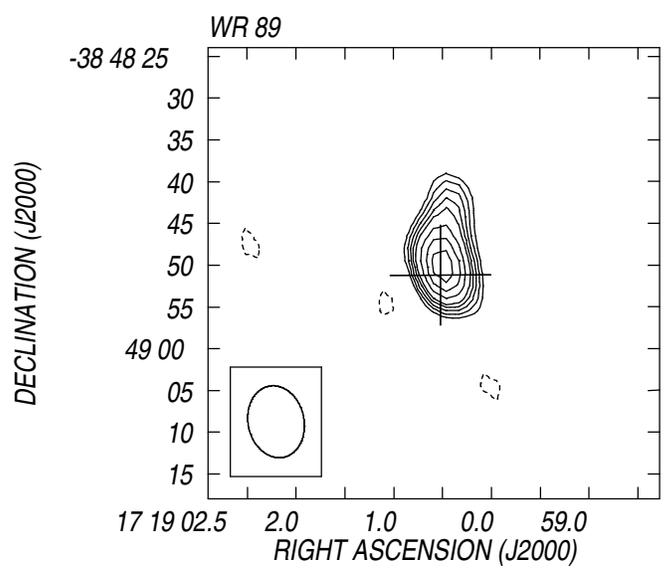
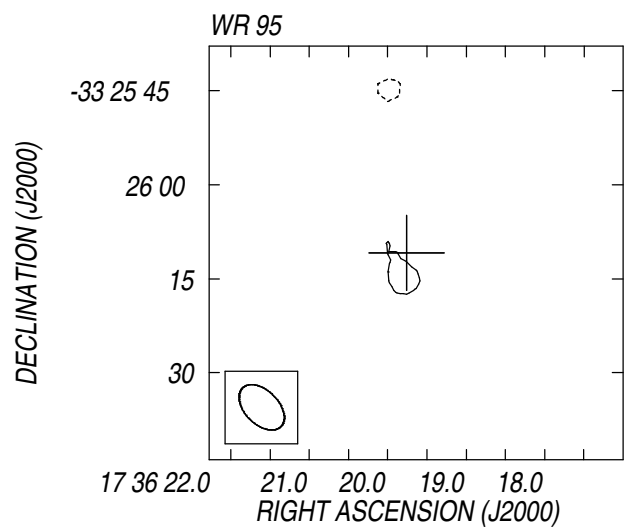
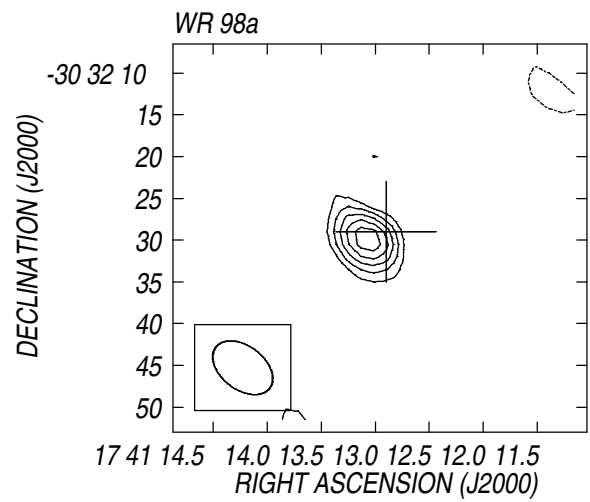
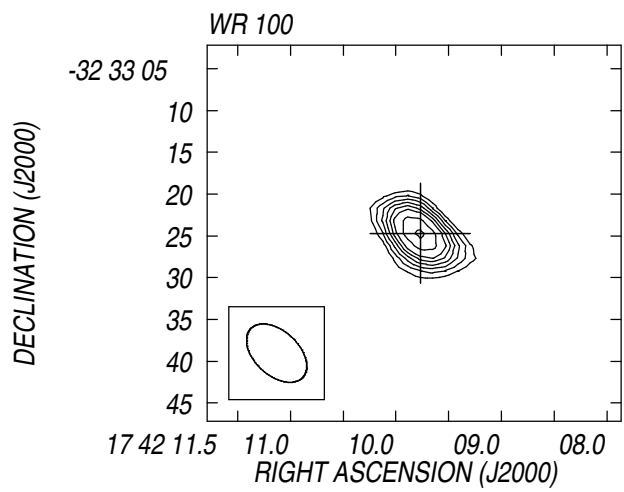
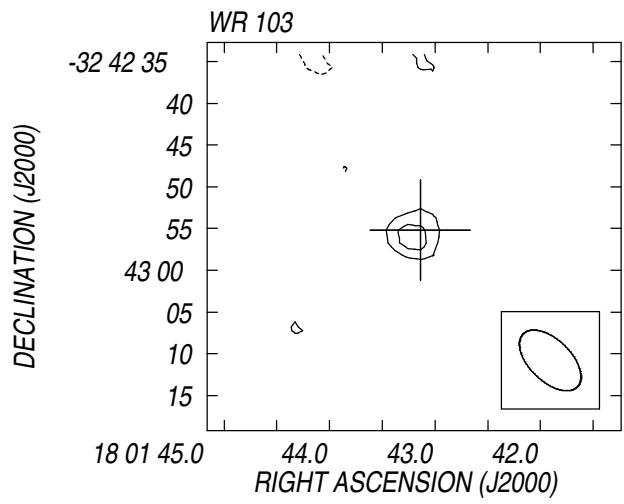
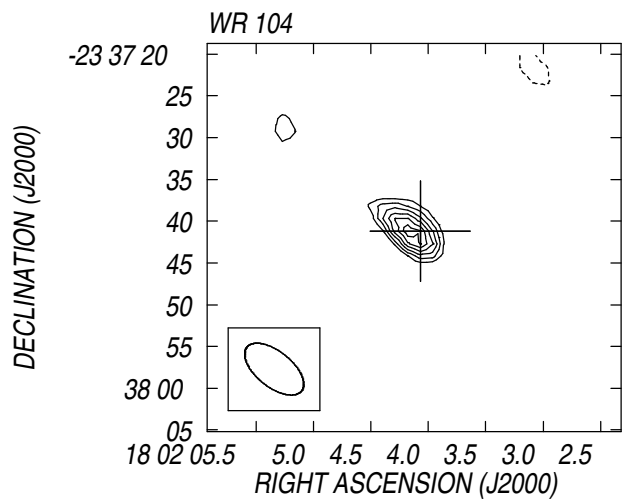
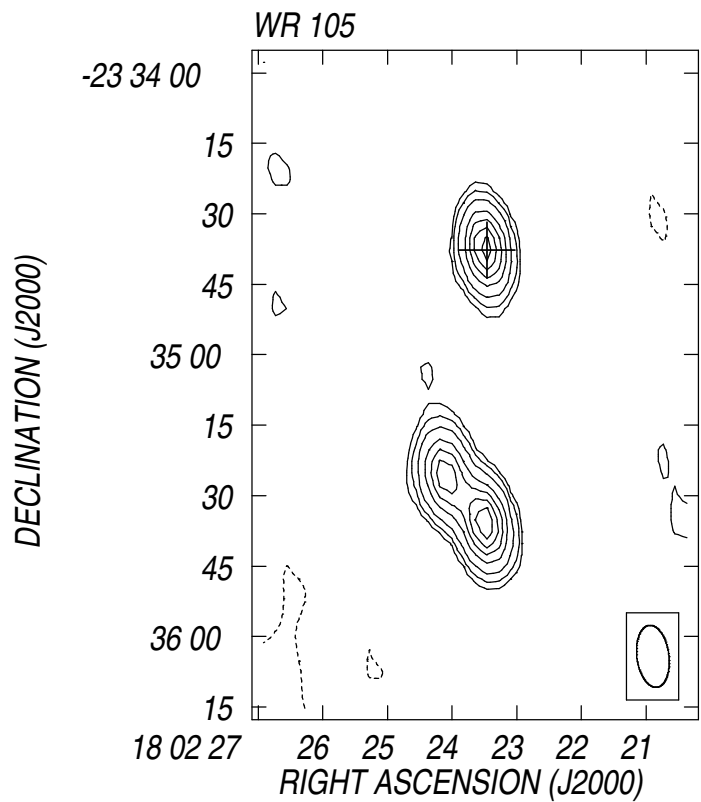
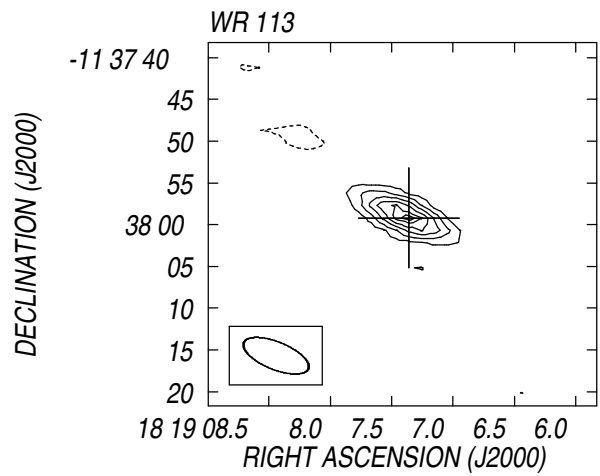
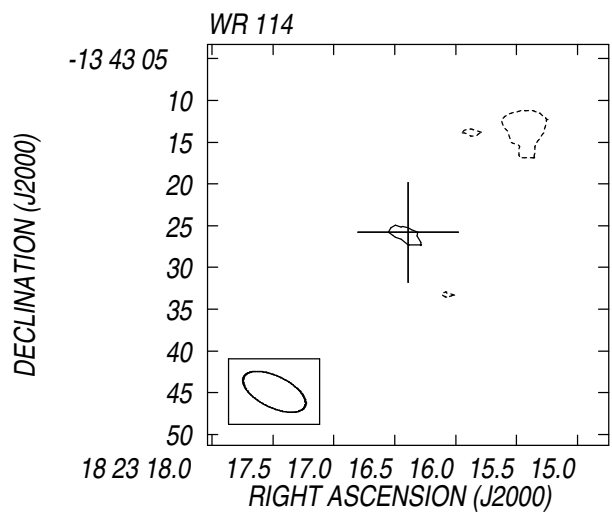
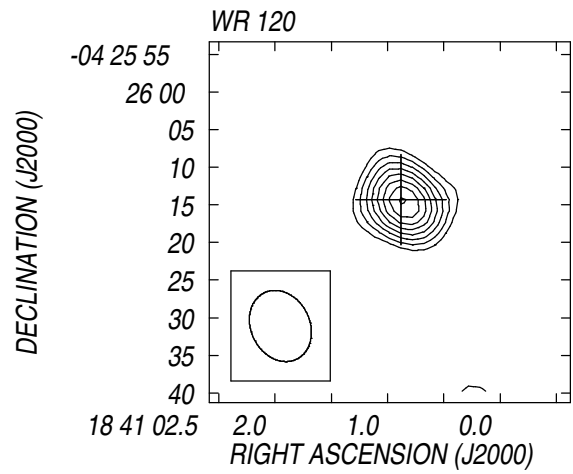
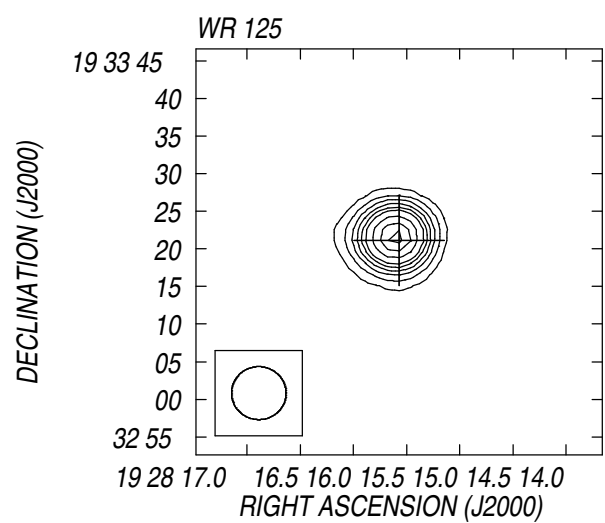
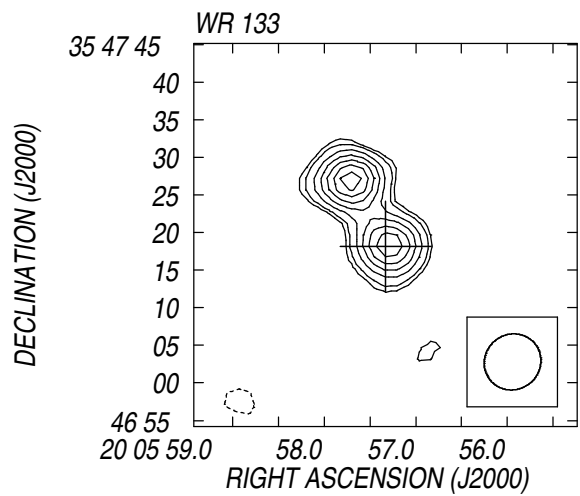
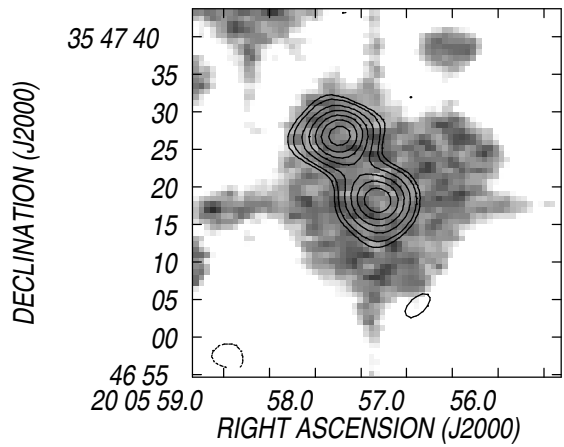
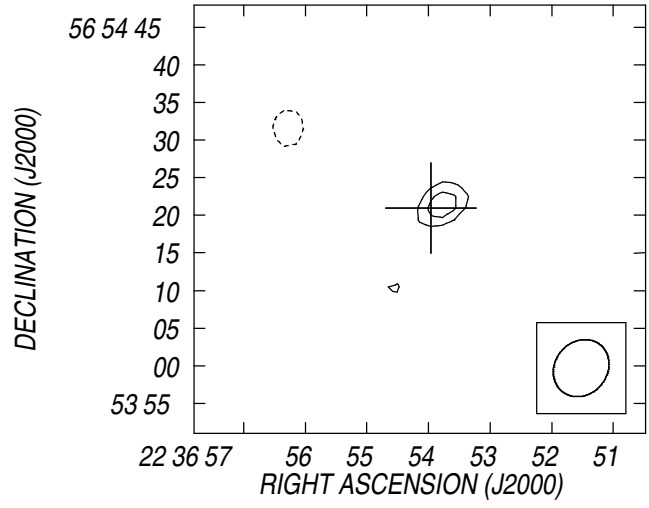
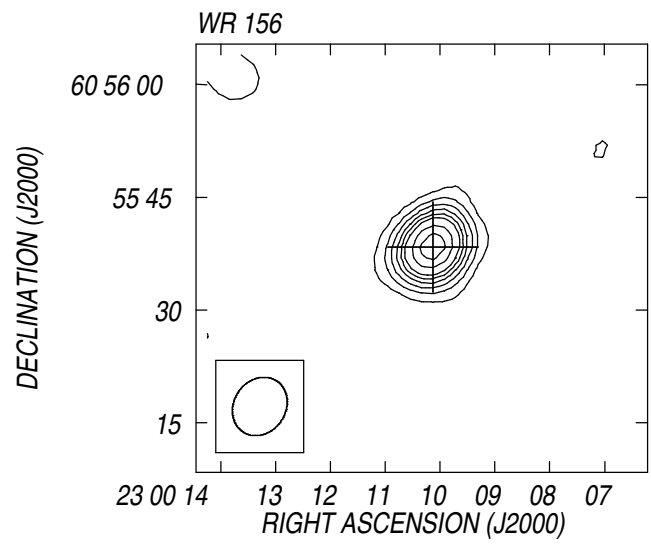
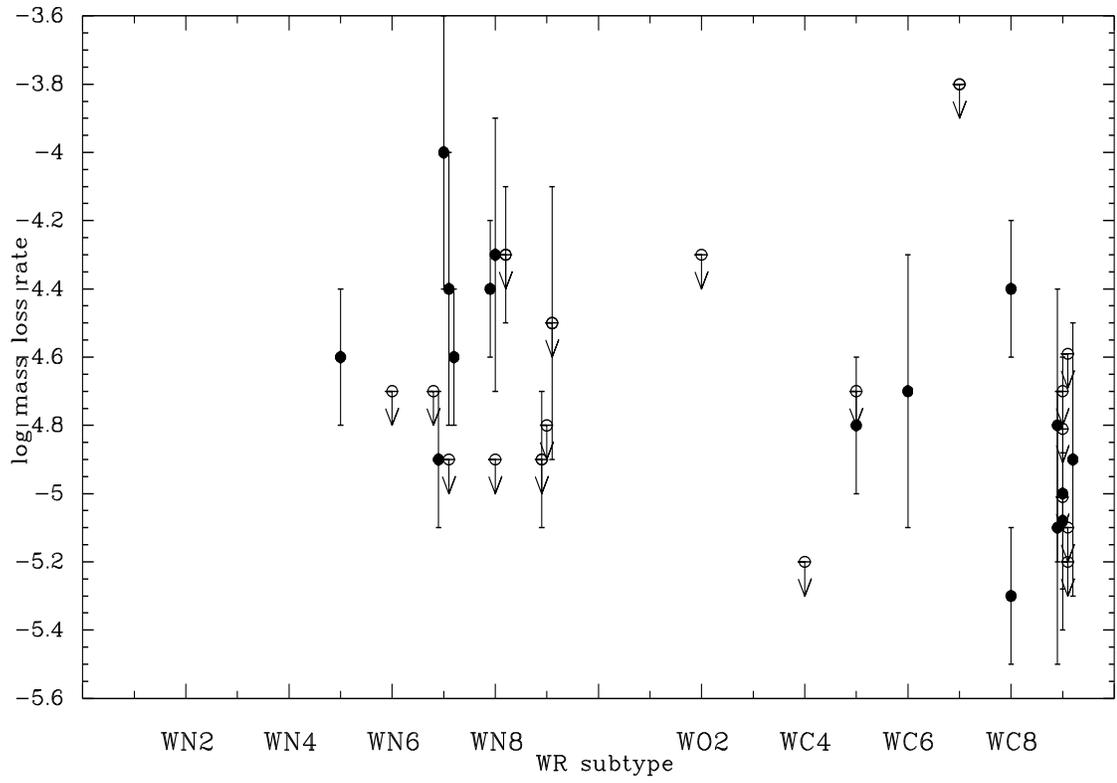
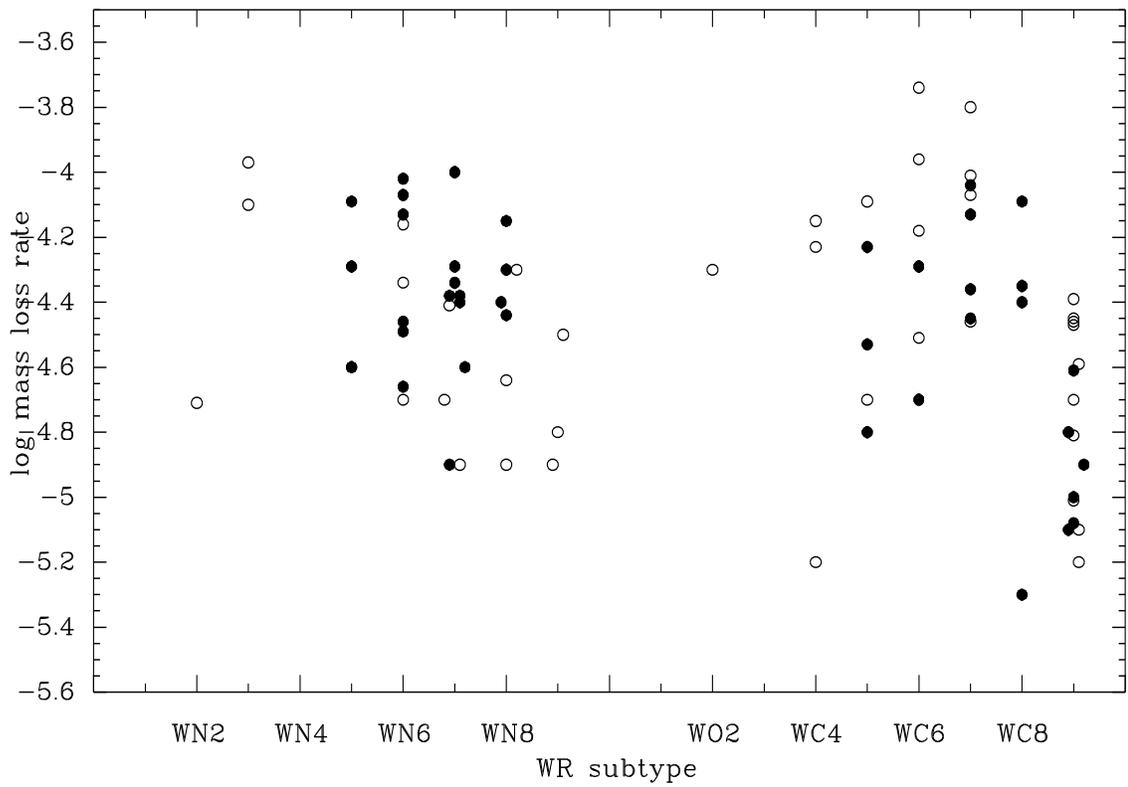
\end{document}